\documentclass[sigconf,nonacm]{acmart} 
\renewcommand\footnotetextcopyrightpermission[1]{}  
\AtBeginDocument{%
  }

\usepackage{booktabs}
\usepackage{xurl}
\usepackage{enumitem}
\usepackage{comment}
\begin{document}

\title{Predicting Abandonment of Open Source Software Projects with An Integrated Feature Framework} 

\author{Yiming Xu}
\orcid{0009-0001-7620-1381}
\affiliation{%
\institution{School of Computer Science, Peking University}
  \institution{Key Laboratory of High Confidence Software Technologies, Ministry of Education}
  \city{Beijing}
  \country{China}
}
\email{xym@pku.org.com}

\author{Runzhi He}
\orcid{}
\affiliation{%
\institution{School of Computer Science, Peking University}
  \institution{Key Laboratory of High Confidence Software Technologies, Ministry of Education}
  \city{Beijing}
  \country{China}
}
\email{rzhe@pku.edu.cn}

\author{Hengzhi Ye}
\orcid{}
\affiliation{%
  \institution{School of Computer Science, Peking University}
  \institution{Key Laboratory of High Confidence Software Technologies, Ministry of Education}
  \city{Beijing}
  \country{China}
}
\email{hzye@stu.pku.edu.cn}

\author{Minghui Zhou}
\orcid{}
\affiliation{%
  \institution{School of Computer Science, Peking University}
  \institution{Key Laboratory of High Confidence Software Technologies, Ministry of Education}
  \city{Beijing}
  \country{China}
}
\email{zhmh@pku.edu.cn}

\author{Huaimin Wang}
\orcid{}
\affiliation{%
  \institution{National University of Defense Technology}
  \institution{National Key Laboratory of Parallel
and Distributed Computing}
  \city{Changsha}
  \state{Hunan}
  \country{China}
}
\email{hmwang@nudt.edu.cn}


\renewcommand{\shortauthors}{Xu et al.}

\begin{abstract}
Open Source Software (OSS) is a cornerstone of contemporary software development, yet the increasing prevalence of OSS project abandonment threatens the quality, security, and resilience of global software supply chains. 
Although previous research has explored a range of abandonment prediction methods---drawing on signals from simple activity metrics (e.g., stars, commits) to social network and behavioral analyses---these methods often demonstrate unsatisfactory predictive performance, further plagued by imprecise abandonment discrimination, limited interpretability, and a lack of large, generalizable datasets.
In this work, we address these challenges by reliably detecting OSS project abandonment through a dual approach: explicit archival status and rigorous semantic analysis of project documentation or description. Leveraging a precise and scalable labeling pipeline, we curate a comprehensive longitudinal dataset of 115,466 GitHub repositories, encompassing 57,733 confirmed abandonment repositories, enriched with detailed, timeline-based behavioral features.
Building on this foundation, we introduce an integrated, multi-perspective feature framework for abandonment prediction, capturing user-centric, maintainer-centric, and project evolution features. Survival analysis using an AFT model yields a high C-index of 0.846, substantially outperforming models confined to surface features. Further, feature ablation and SHAP analyses confirm both the predictive power and interpretability of our approach.
We further demonstrate practical deployment of a GBSA classifier for predicting package risk in the openEuler ecosystem.
By unifying precise labeling, multi-perspective features, and interpretable modeling, our work provides reproducible, scalable, and practitioner-oriented support for understanding and managing abandonment risk in large OSS ecosystems. Our tool not only predicts abandonment but also enhances program comprehension by providing actionable insights into the health and sustainability of OSS projects.

\end{abstract}






\maketitle

\section{Introduction}
Open-source software (OSS) is the foundation of modern software engineering, powering diverse applications and enterprise systems. Collaborative platforms such as GitHub have enabled developers to reuse third-party OSS components at an unprecedented scale, significantly accelerating development and reducing cost. According to the 2025 Open Source Security and Risk Analysis (OSSRA) report~\cite{OSSRA2025}, over 97\% of analyzed software projects across industries contain OSS dependencies, underscoring the centrality of OSS in contemporary software supply chains.

Meanwhile, the growing dependence on OSS means the growing risk of ecosystem-wide disruptions.
OSS projects are typically maintained by a small number of developers on a voluntary basis. These developers may discontinue their involvement in project maintenance for various reasons, such as the project no longer meeting public needs, being replaced by competing alternatives, loss of personal interest, and so forth~\cite{10.1145/3106237.3106246}.
Once no one continues to maintain the project, the OSS is considered abandoned, which is not rare in the open source ecosystem, and it poses sudden risks of software defects, security vulnerabilities, and more to numerous downstream users and dependent software systems~\cite{DBLP:journals/corr/abs-1709-04638}.
The OSSRA report finds that 91\% of audited OSS components exhibited no clear signs of maintenance in the past two years~\cite{OSSRA2025}, highlighting the prevalence of this issue. 
Notably, high-profile OSS projects may also run into abandonment---well-known examples include \emph{Atom}~\cite{atom}, \emph{Brackets}~\cite{brackets}, and \emph{faker.js}~\cite{faker.js}.

Predicting OSS project abandonment is crucial not only for risk mitigation, but also for supporting practical program comprehension tasks, such as dependency analysis and architectural assessment. Interpretable risk signals allow developers to better understand and manage the sustainability and evolution of their software dependencies, leading to more robust and informed decision-making. 
While extensive research has explored the potential of data-driven abandonment prediction~\cite{10.1016/j.infsof.2010.05.001, Xia2022, HSPM}, two challenges must be addressed for these prediction techniques to be practical. 

The first challenge lies in the discrimination of abandonment.
Some previous studies have defined project abandonment solely based on activity metrics. They often assume a project is abandoned after a fixed threshold of inactivity~\cite{9796257, 2025arXiv250418971M}. 
We found counter-examples of ``revival'' in real-life non-trivial OSS projects to their assumption~\cite{supervisor, backbone, underscore.string}, indicating that inactivity cannot definitively determine abandonment. Several prior works have employed explicit abandonment statements as the criterion to discriminate abandonment, and our method also adopts this approach for determination.
Notably, previous studies~\cite{HSPM} often used direct keyword matching in project descriptions for discrimination. Within our manually labeled set of 1,174 randomly selected keyword-matched OSS repositories, only 404 (34.4\%) had README files or descriptions that genuinely indicated the repository was abandoned. The remaining 768 repositories (65.6\%) contained no explicit statement of abandonment. Both SetFit and LLM methods, and prior research~\cite{10.1145/3106237.3106246} yielded similar results, indicating that the previously used keyword-matching method for identifying abandonment contained a substantial proportion (over 60\%) of false positives---a figure that cannot be overlooked. These repositories, lacking genuine abandonment statements, likely triggered the keyword matches for various reasons, such as declaring a specific version abandoned, stating that another project was abandoned, or indicating that a particular component was abandoned.
Some studies~\cite{10.1145/3106237.3106246,miller2023,miller2025} have used manual labelling for accurate identification of abandonment statements in text, albeit often on a small scale due to the intensive labor involved.

The second challenge concerns predictive features.
Existing approaches typically use surface features such as the number of \textit{star}, \textit{commit}, or \textit{pull request (PR)}, which are volatile, susceptible to manipulation~\cite{DBLP:journals/corr/abs-2412-13459}, and insufficient for capturing deeper sustainability dynamics such as governance transitions, contributor attrition, or changes in user engagement. 
Other approaches that incorporate more high-level features and advanced methods often lack interpretability, such as straightforward feeding information into GNNs or other complex models~\cite{HSPM}. These features fail to support genuine program comprehension regarding a project's health and sustainability. Often, they merely produce a numeric value whose semantic meaning is difficult to interpret and actionable insights for maintainers remain unclear~\cite{GNN-explanation}. Regarding some of the more interpretable advanced features proposed in existing studies, certain limitations remain; we provide a detailed discussion in Section~\ref{sec:feature-framework}.

Project abandonment is identified through a union of two criteria: (1) explicit GitHub \emph{archived} status~\cite{github-archiving, github-archiving-doc}, and (2) unambiguous abandonment statements in project documentation or description~\cite{10.1145/3106237.3106246}.
We implemented this through a precise pipeline that combines manual labeling, ML, and LLM.
This approach virtually eliminates the false positives, which accounted for over 60\% of labels in prior keyword-matching methods, while enabling rapid processing of large data volumes to construct a large-scale dataset.

From this precise labelling, we have generated a large-scale longitudinal dataset of 115,466 GitHub repositories, including 57,733 with validated abandonment events and comprehensive behavioral timelines (Section~\ref{subsec:dataset}).
We then propose a multi-dimensional, interpretable feature framework that goes beyond surface features by jointly modeling: \textbf{(1)} user influence via a time-aware, weighted bipartite user-project network; \textbf{(2)} maintainer behavior, including activity recency and response latency; and \textbf{(3)} project evolution characteristics, such as community role balance and feature/bugfix contribution shifts (Section~\ref{subsec:feature-design}).
Using an accelerated failure time (AFT) model, our approach achieves a high C-index (0.846), filling a critical gap in existing research while providing superior predictive power, interpretability, and early warning capabilities.
Beyond predictive accuracy, our feature framework enables developers and ecosystem stakeholders to better comprehend the underlying causes, behavioral signals, and community dynamics of OSS project abandonment. 
Furthermore, to demonstrate real-world applicability, we deployed a GBSA classifier trained on our framework within the openEuler ecosystem to conduct proactive risk screening of its packages (Section~\ref{sec:framework-application}). This demonstrates the framework's practical utility in assessing risks for large-scale, real-world software supply chains.

To sum up, in this paper, we: \textbf{(1)} Proposed a precise method for discrimination true abandonment; \textbf{(2)} Constructed a large-scale, longitudinally labeled dataset of OSS repositories; \textbf{(3)} Designed a multi-dimensional, interpretable feature framework for predicting abandonment; \textbf{(4)} Empirically demonstrated superior predictive performance; \textbf{(5)} Validated the real-world utility of our approach through deployment in the openEuler ecosystem.

\begin{figure*}[ht] 
    \centering
    \includegraphics[width=0.9\linewidth]{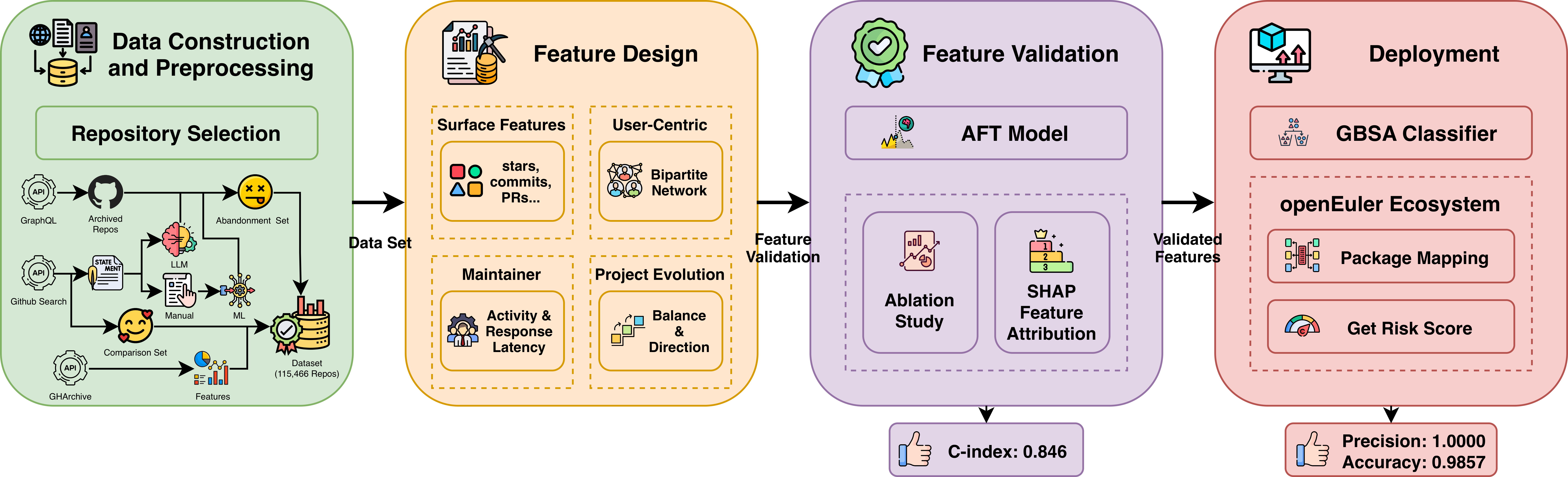}
    \vspace{-0.5em}
    \caption{Methodology Overview.}
    \vspace{-0.5em}
    \label{fig:method-overview}
\end{figure*}

By establishing a rich foundation for program comprehension and risk awareness, our work contributes to software supply chain risk assessment, OSS dependency management, and sustainable ecosystem governance. As illustrated in Figure~\ref{fig:method-overview}, this foundation is built upon a methodological framework comprising four core components, which are tightly integrated to ensure scientific rigor, feature interpretability, and real-world applicability.


\section{Background}
\label{sec:background}

Predicting open source project abandonment has attracted significant research interest in software engineering. Early work primarily adopted survival analysis methods from reliability engineering, using models like Cox proportional hazards and Kaplan-Meier estimators to examine how factors such as active contributor count and major version releases correlate with project survival~\cite{10.1016/j.infsof.2010.05.001, 9796257, Khondhu2013IsIA, Samoladas2010SurvivalAO, 9743934, 10.1145/3236024.3236062, 9631870}.

With the growth of social coding platforms, recent studies have shifted toward machine learning methods. Most define abandonment through symptoms like prolonged inactivity in commits or pull requests, or occasionally through maintainer declarations in documentation~\cite{HSPM, 9588891, 10.1109/MS.2022.3163011,miller2023,miller2025}. Feature engineering in these works relies heavily on surface-level activity counts—stars, commits, issues, PRs—sometimes supplemented with social or network features, or advanced architectures such as GNNs and ensemble methods~\cite{10.1007/978-981-15-7984-4_20, HSPM, Xia2022, 10.1145/3239235.3240501, COELHO2020106274}.

Despite these advances, several key limitations hamper the robustness, interpretability, and practical adoption of the current state-of-the-art:

\textit{Imprecise Abandonment Criteria.} The absence of precise abandonment criteria in most previous work leads to high false-positive rates. For example, keyword-matching mislabels over 60\% of repositories, and definitive signals like GitHub's "archived" status are often overlooked.

\textit{Over-reliance on surface features.}
Most prior approaches rely primarily on surface features (e.g., \emph{stars}, \emph{commits}, \emph{PRs}, \emph{issues}), which, despite being easy to obtain, may fail to capture deeper signals of project sustainability. The limitations of such surface features are further discussed in Section~\ref{subsubsec:surface-features}.

\textit{Opaque feature semantics in advanced models.}
Some studies attempt to improve prediction by extending surface features with network-based or behavioral signals, or by employing complex models~\cite{10.1007/978-981-15-7984-4_20, HSPM, Xia2022}. However, these approaches often yield limited predictive performance and lack interpretable explanations of how such features contribute to outcomes, relying instead on model capacity alone. This opacity limits their practical value to ecosystem maintainers who need transparent, trustworthy insights for decision-making.

\textit{Limited data scale and ecosystem coverage.}
Existing datasets are often small, highly curated, or biased towards repositories with high visibility (e.g., high star counts), and as such do not represent the diversity of OSS projects. As a consequence, findings from these studies may lack generalizability to the broader software ecosystem.

\textit{Insufficient empirical validation and practical deployment.}
Even as machine learning techniques become more sophisticated, their practical effectiveness is rarely examined at ecosystem scale. Most approaches are evaluated only in isolated, research-focused settings and are not validated within actual package management or release governance workflows.

While recent efforts address some gaps, challenges remain in scaling interpretable methods, improving abandonment discrimination, and enabling actionable risk assessment in production environments.

\section{Foundations for Prediction}
\label{sec:methodology}

This section details the foundational methodology for our study on OSS project abandonment prediction. We first address the critical prerequisite of operationalizing abandonment through a precise, dual-criteria definition. We then describe the construction of the large-scale, longitudinal dataset that underpins all subsequent analysis and modeling in this work.

\subsection{Operationalization of Abandonment}
\label{subsec:operationalization-of-abandonment}

Accurately predicting OSS project abandonment is critical yet challenging. A prerequisite for accurate prediction is the precise identification of abandonment events.
Existing research shows inconsistent operationalizations: some rely solely on prolonged inactivity (e.g., no \emph{commits} within a fixed period)~\cite{HSPM,9796257,2025arXiv250418971M}, while others use keyword matching to detect maintainer declarations~\cite{10.1145/3106237.3106246, HSPM}.
Unfortunately, keyword-based methods incur high false-positive rates, as confirmed in our study, while precise manual labeling does not scale to large datasets.

In this work, we define a project as abandoned if at least one of the following conditions is met:
\begin{enumerate}
    \item \textbf{Explicit archival status:} The repository is marked as ``archived'' on GitHub, making it read-only and signaling explicit maintainer abandonment~\cite{github-archiving-doc,github-archiving}. The archiving timestamp defines the abandonment time.
    \item \textbf{Semantic statement of abandonment:} The project's top-level documentation contains clear textual statements (e.g., “abandoned,” “no longer maintained,” “deprecated,” or “end-of-life”). We detect these using a hybrid pipeline combining keyword filtering, manual verification, machine learning and LLM-based classification, as detailed in Section~\ref{subsec:dataset}.
\end{enumerate}

This dual-criterion approach ensures both recall (capturing text-based signals beyond archiving) and precision (eliminating over 60\% of false positives from keyword matching alone). The resulting high-quality labels underpin all subsequent predictive modeling in this work.

    
    
    


\subsection{Data Construction and Preprocessing}
\label{subsec:dataset}

To support robust and generalizable prediction, we constructed a comprehensive dataset capturing longitudinal behavioral traces and rigorously validated abandonment labels for a large population of OSS repositories. The construction process included repository selection, multi-source data collection, careful labeling, temporal aggregation, and quality control, as outlined in the \emph{Data Construction and Preprocessing} box of Figure~\ref{fig:method-overview}.

\paragraph{Repository Selection and Scope.}

We focus on non-fork repositories hosted on GitHub~\cite{github}, the world's largest collaborative development platform. To balance representativeness and tractability, we include all repositories created between 2011 and 2025 with at least \emph{32 stars}, following prior studies~\cite{xwx} and API limit~\cite{graphql} considerations. This threshold filters out trivial or inactive repositories, while preserving diversity in languages, domains, and activity levels.

\paragraph{Ground Truth Labeling.}
We implement a three-step pipeline to produce high-quality, reproducible abandonment labels:

\begin{enumerate}
    \item \textbf{Archived Repository Collection.}
    Using the GitHub \\ GraphQL API~\cite{graphql}, we collect repositories explicitly marked as "archived" --- a read-only state treated as a strong, platform-verified abandonment signal. The archival timestamp is recorded as the abandonment date.
    \item \textbf{Abandonment Declaration Mining.}  
    Many developers choose to declare abandonment in README files or repository descriptions rather than use the archive flag~\cite{10.1145/3106237.3106246,miller2023,miller2025}. We identified such cases using a semantic statement extraction strategy:
    (1) Following prior works~\cite{HSPM,10.1145/3106237.3106246,miller2025}, we first use a list of abandonment keywords (e.g., ``abandoned'', ``deprecated'', ``unmaintained'', etc.) is first used to retrieve candidate repositories. To mitigate false positives due to ambiguous usage (e.g., abandon a dependency rather than the project itself), we perform manual annotation on 1,174 (3\%) randomly selected samples.
    (2) The annotation task is conducted using Label Studio~\cite{labelstudio}, where two authors label each sample based on textual signals. The first round of labeling ended with an agreement of 95\% and a Cohen’s kappa of 0.814, which suggests a strong agreement. All inconsistencies
    were resolved through discussion, leading to a final consensus on the entire dataset.
    (3) Using 587 samples for training and 587 for validation, we fine-tune a SetFit classifier~\cite{2022arXiv220911055T} based on sentence transformers~\cite{reimers-gurevych-2019-sentence}. Our best model, built on the paraphrase-mpnet-base-v2 checkpoint~\cite{paraphrase-mpnet-base-v2}, achieved 0.96 accuracy, 0.96 recall, and 0.90 precision.
    \item \textbf{LLM-Assisted Labeling.}  
    To scale high-precision labeling, we evaluated multiple large language models (LLMs) on the 3\%-sample manual labelling benchmark. Among the tested models (including GPT-4o, DeepSeek-V3, DeepSeek-R1, Claude-3.7, and Qwen series), GPT-4o performed best with accuracy, precision, recall, and F1 achieving 0.9463, 0.9429, 0.8985, and 0.9202. This model and SetFit was consequently selected to label remaining samples. 
    We observed 95.91\% agreement between GPT-4o and our SetFit classifier. All disagreements were resolved through discussion, resulting in 1,515 relabeled samples during manual cross-validation.
\end{enumerate}

This process yields 57,733 repositories with confidently labeled abandonment dates from archival flags or textual declarations, eliminating all the 60\%+ false positive rate observed in prior keyword-based methods.

\paragraph{Longitudinal Behavioral Data Extraction.}
To capture project evolution, we aggregate time-series repository activities using \\ GHArchive~\cite{gharchive2024} and GitHub API. Monthly records include: 
\begin{itemize}
    \item Code contributions: \emph{commits} (including lines changed), \emph{PR creations}.
    \item Community engagement: \emph{issues}, \emph{comments}, \emph{stars}.
    \item Metadata changes: \emph{tags}, description or README updates.
\end{itemize}
Compared with other public datasets (e.g., GHTorrent~\cite{gousios2013}), \\ GHArchive is leveraged for its completeness and efficiency in reconstructing historical events at scale. Events are parsed via distributed ETL and stored in ClickHouse for efficient querying.

\paragraph{Quality Control and Initial Analysis.}

Our quality control process began with the removal of repositories exhibiting inconsistent, corrupted, or incomplete event histories, typically resulting from missing API data or API reconciliation failures. This initial filtering step affected a negligible portion (0.3\%) of the dataset, ensuring data integrity without materially influencing the overall results.

\begin{figure}[H]
    \centering
    \includegraphics[width=1\linewidth]{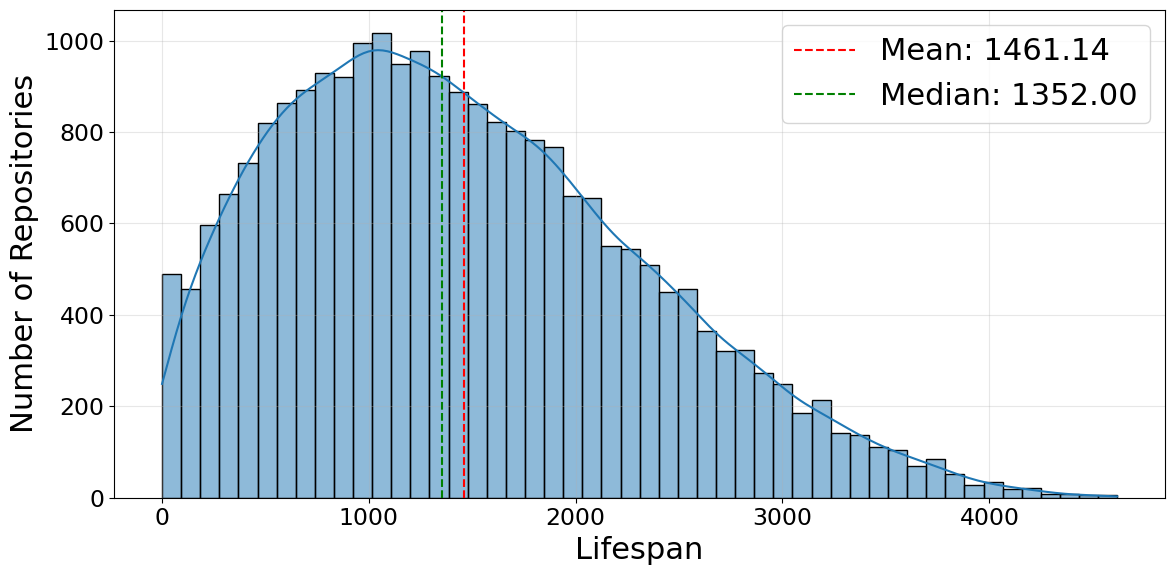}
    \vspace{-2em}
    \caption{Lifespan distribution of abandoned repositories.}
    \label{fig:lifespan-distribution}
    \vspace{-0.5em}
\end{figure}

Initial statistical analysis compared star counts and lifespans between active and abandoned repositories to validate dataset representativeness and screen for potential annotation biases.
Figure~\ref{fig:lifespan-distribution} shows the lifespan distribution of 23,730 abandoned projects,  averaging nearly four years with high variability---underscoring the difficulty of predicting abandonment.

Further insights are drawn from the star count distributions in Table~\ref{tab:star-distribution}. Abandoned projects demonstrate significantly lower mean and maximum star counts compared to the full population, consistent with the expectation that abandoned projects attract diminishing attention. The considerably smaller standard deviation and maximum star counts further indicates that abandoned projects lack the highly popular repositories found in the overall dataset.

Notably, abandoned projects show a higher median star count. Abandonment is discriminated through explicit maintainer actions, which are more likely to occur in projects that have already garnered a moderate level of community attention. Consequently, repositories abandoned silently with minimal user engagement are under-represented in our labeled abandoned cohort, slightly inflating the median star count for confirmed abandoned projects. This initial analysis validates our data quality and proves that our data is representative.

\begin{table}[h]
    \centering
    \caption{Distribution of star counts in the dataset.}
    \vspace{-1em}
    \begin{tabular}{lccccc}
    \toprule
    \textbf{Category}   &  \textbf{Mean} &  \textbf{Std} &  \textbf{Min} &    \textbf{50\%} &   \textbf{Max} \\ 
    \midrule
    All Projects   & 382.88 & 2241.63 & 32 & 83 & 222257 \\ 
    Abandoned Projects   & 268.79 & 443.33 & 32 & 126 & 2383 \\ 
    \bottomrule
    \end{tabular}
    
    \label{tab:star-distribution}
    \vspace{-1em}
\end{table}

\section{Feature Framework}
\label{sec:feature-framework}
Numerous factors contribute to the abandonment of OSS projects~\cite{10.1145/2393596.2393662,7816485}. To model this phenomenon comprehensively, we design a multi-perspective feature framework. Our framework integrates several feature categories, beginning with surface activity features which establish a baseline using common features (e.g., \emph{stars}, \emph{commits}) widely used in prior work. Moving beyond this foundation, we introduce three novel feature groups that constitute our primary contribution: (1) \textbf{user-centric features}---to our best knowledge, the first of their kind proposed for abandonment prediction, modeling the OSS ecosystem as a weighted bipartite graph to capture the structure and evolution of user-repository interactions; (2) \textbf{maintainer-centric features}, whose novelty lies in explicitly identifying maintainers---a critical distinction from generic contributors often glossed over in studies---leveraging their specific engagement and responsiveness signals for prediction; and (3) \textbf{project evolution features}, which mark another first by operationalizing into a predictive model deeper lifecycle characteristics such as the changing collaboration structure, development focus (e.g., the bug-fix ratio~\cite{10.1145/2901739.2903495,10.1145/3340544}), and periodic behavioral patterns. 
In the following subsections, we delineate the design and computation for each feature group and explain how each captures behavioral mechanisms relevant to program comprehension, with empirical validation provided in Section~\ref{subsec:feature-test}.
We next describe each group and their relevance to program comprehension in Section~\ref{subsec:feature-design}, with empirical tests in Section~\ref{subsec:feature-test}.

\begin{table*}[h]
    \centering
    \caption{Categorized list of features used in the abandonment prediction framework.}
    \begin{tabular}{lp{10.5cm}}
        \toprule
        \textbf{Feature} & \textbf{Description} \\
        \midrule
        \multicolumn{2}{l}{\textbf{Surface Activity}} \\
        \texttt{stars} & Number of stars received by the repository. \\
        \texttt{commits} & Number of commits in the repository. \\
        \texttt{issues} & Number of issues opened in the repository. \\
        \texttt{prs} & Number of pull requests submitted. \\
        \texttt{tags} & Number of tags in the repository. \\
        \texttt{comments} & Number of comments made on issues and PRs. \\

        \midrule
        \multicolumn{2}{l}{\textbf{User-Centric}} \\
        \texttt{weight} & User-centric feature (see Section~\ref{subsubsec:weight}). \\
        \texttt{weight\_rank\_pct} & Percentile rank of \texttt{weight}. \\
        \texttt{weight\_zscore} & Z-score normalized value of \texttt{weight}. \\

        \midrule
        \multicolumn{2}{l}{\textbf{Maintainer-Centric}} \\
        \texttt{latest\_maintainer\_activity\_interval} & Time interval since the last maintainer action. \\
        \texttt{avg\_response\_time} & Average maintainer response time to issues and PRs. \\
        \texttt{response\_decay\_trend} & Trend of \texttt{avg\_response\_time}; positive indicates degradation. \\

        \midrule
        \multicolumn{2}{l}{\textbf{Project Evolution} (By default, data within the recent 6 months)} \\
        \texttt{maintainer\_contrib\_ratio} & Proportion of maintainer activities over all repository activities. \\
        \texttt{contrib\_diversity} & Contributor diversity measured by the 1-Gini index. \\
        \texttt{balance\_index} & Structural balance combining maintainer and others' contribution share. \\
        \texttt{activity\_deviation} & Normalized deviation of recent 3-month activity from historical mean. \\
        \texttt{quarterly\_deviation} & Deviation from periodic quarterly activity trends. \\
        \texttt{feature\_ratio} & Ratio of feature-related PRs among all PRs. \\
        \texttt{bugfix\_ratio} & Ratio of bug-fixing PRs among all PRs. \\
        \texttt{bugfix\_feature\_ratio} & Ratio of bugfix PRs to feature PRs, indicating maturity shift. \\

        \bottomrule
    \end{tabular}
    
    \label{tab:feature-categories}
\end{table*}

\subsection{Feature Design}
\label{subsec:feature-design}

We begin by summarizing surface features commonly used in existing work, then introduce our enhanced (new) feature groups. Table~\ref{tab:feature-categories} provides an overview of all features in our framework.

\subsubsection{Surface Features}
\label{subsubsec:surface-features}

Prior work mainly uses surface features: counts of \emph{stars}, \emph{commits}, \emph{issues}, \emph{PRs}, etc., all of which are directly accessible via public platform APIs and have intuitive appeal as signals of project activity~\cite{HSPM, 9588891, 9796257, 10.1145/2627508.2627515, 10.1145/3106237.3106246, 10.1145/2972958.2972966}. 

While intuitive, these features suffer from key limitations:
\textbf{(1) No user differentiation}: all interactions are weighted equally, overlooking differences in user roles. For example, a star from a renowned developer is more meaningful than one from a novice, but surface features flatten such distinctions;
\textbf{(2) Vulnerability to manipulation}: they can be artificially inflated (e.g., via paid \emph{stars}~\cite{DBLP:journals/corr/abs-2412-13459}), undermining their reliability for predicting project health;
\textbf{(3) Limited expressiveness}:  they fail to capture community roles, collaboration patterns, and project evolution.

To address these shortcomings from a program comprehension perspective, we propose three structured feature categories that move beyond surface counts to capture interpretable behavioral and structural signals.

\subsubsection{User-Centric Features}
\label{subsubsec:weight}

\begin{figure}[H]
    \centering
    \includegraphics[width=1\linewidth]{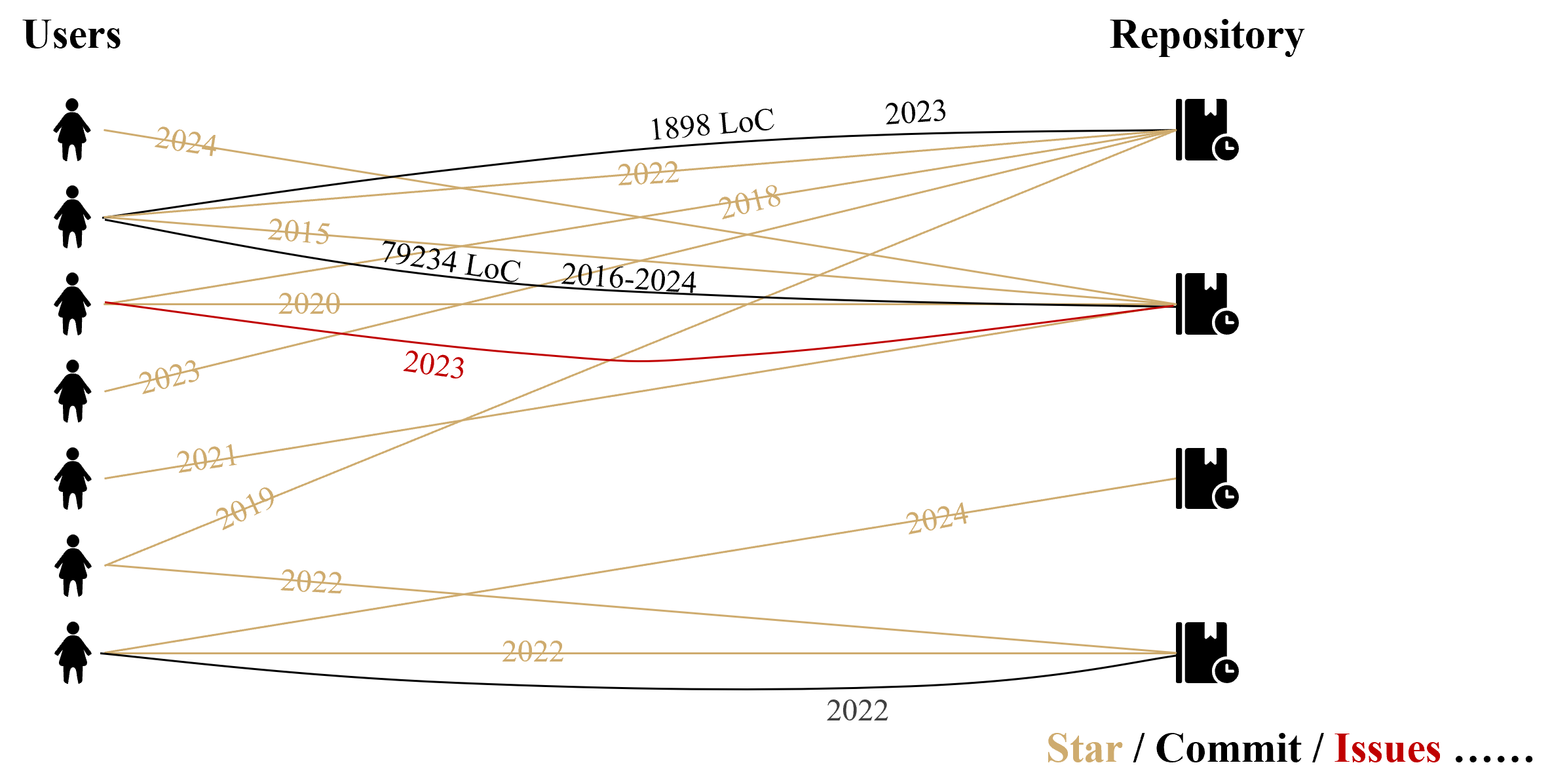}
    \caption{Illustration of the user-repository bipartite graph.}
    \label{fig:bi-graph}
    \vspace{-1em}
\end{figure}

We introduce user-centric features based on the structure and evolution of user-repository interactions---a novel perspective not previously explored in abandonment prediction. Our key insight is that users attracted to high-quality projects tend to be more experienced, and repositories that engage these users are more likely to be sustainable.
Some studies have uncovered new information within networks~\cite{community-structure,10.1145/1453101.1453105,5635057}. We represent the OSS ecosystem as a weighted bipartite graph~\cite{BLINCOE201630} (Figure~\ref{fig:bi-graph}), where nodes represent users or repositories and timestamped edges capture their interactions. Figure~\ref{fig:bi-graph} serves as a schematic of this structure, and details are provided in this subsection.
This approach enables us to capture complex patterns of user influence and project health that surface features alone cannot reveal.

\paragraph{Interaction-Based Influence Modeling.}
We model user-repository relationships through four interaction types: \emph{starring}, \emph{committing}, \emph{forking}, and \emph{issues}. Reflecting that engagement quality and timing better indicate sustainability than volume (supported by the superior C-index of S+U over S+H in Table~\ref{tab:eval-model}), we assign base weights to each interaction type. Given the dataset's size, which precludes exhaustive tuning, we determined these weights ($w_{orig}^{star} = 1$, $w_{orig}^{commit} = 8$, $w_{orig}^{fork} = 4$, $w_{orig}^{issue} = 2$) by evaluating multiple combinations on a 500-repository subset. This configuration performed well on the full dataset, and further parameter optimization is secondary to our primary contribution: establishing the overall effectiveness of user-centric features for abandonment prediction and understanding the status of OSS projects.
To prioritize early or sustained engagement, we apply a temporal decay factor to each interaction, defined as:

\begin{equation}
\begin{aligned}
    pos_{(up)}^t = \frac{k_{(up)}^t}{N_p^{t'}} , \quad
    D(pos) = \frac{1}{1 + p} , \quad
    d_{(up)}^t = D(pos_{(up)}^t)
\end{aligned}
\end{equation}

Here, $pos_{(up)}^t$ denotes the relative position of the interaction between user $u$ and OSS project (repository) $p$ at time $t$ with second-level granularity; $k_{(up)}^t$ indicates that the action is the $k$-th occurrence between $u$ and $p$ in their timeline; and $N_p^{t'}$ is the total number of actions associated with $p$ at the target time for calculating the data. $D(pos)$ is a generic decay function that can be instantiated as needed, and $d_{(up)}^t$ represents the temporal decay factor for the action at $(u, p, t)$, assigning higher weights to earlier interactions to reflect the user's foresight.

\paragraph{Computation of Interaction Weights.}
For each user-repository edge, interaction weights are computed as follows:

\begin{equation}
\begin{aligned}
    &w_{(up)}^{star} = w_{orig}^{star} \times d_{(up)}^t , \quad
    w_{(up)}^{fork} = \sum_{fork_{(up)}} w_{orig}^{fork} \times d_{(up)}^t , \\
    &w_{(up)}^{issue} = \sum_{issue_{(up)}} w_{orig}^{issue} \times d_{(up)}^t
\end{aligned}
\end{equation}

To better quantify the technical contribution of \emph{commits}, we additionally weight each \emph{commit} by the logarithm of lines of code changed $LOC$, a metric derived from commit payloads. Due to API limitations, $LOC$ excludes non-file changes, but we argue this has minimal impact on the user-repository network, as active users typically also make file modifications.

\begin{equation}
    w_{commit} = \sum_{commits_{(up)}} w_{commit}^{orig} \times \lg LOC ~ \textbf{if} ~ LOC>0 ~ \textbf{else} ~ 0
    \label{equa:weight-commit}
\end{equation}

Total user-repository weights are then aggregated:

\begin{equation}
\begin{aligned}
    w_{(up)} = \sum_{o \in Set_{op}} w_{(up)}^{o} , \quad
    {Set_{op}} = \{star, commit, fork, issue \}
\end{aligned}
\end{equation}

\paragraph{Propagation of Influence: HITS-Style Approach.}
Inspired by HITS (Hyperlink-Induced Topic Search)~\cite{10.1145/324133.324140, Prajapati2012ASP}, and leveraging the bipartite graph structure, we employ an iterative algorithm to propagate both influence and project quality as follows:

\begin{equation}
\begin{aligned}
    PQS(p)=\sum_{u\in U_p} w_{(up)}\times UIS(u) , 
    PQS(p)=\frac{PQS(p)}{\sum_{p^{'}\in P}PQS(p^{'})}
\end{aligned}
\end{equation}

\begin{equation}
\begin{aligned}
    UIS(u)=\sum_{p\in P_u} w_{(up)}\times PQS(p) , 
    UIS(u)=\frac{UIS(u)}{\sum_{u^{'}\in U}UIS(u^{'})}
\end{aligned}
\end{equation}

Here, $PQS(p)$ and $UIS(u)$ represent the project quality score for repository $p$ and the user influence score for user $u$, respectively. Scores are initialized to 1 and updated iteratively until convergence. Normalization ensures that scores are comparable among all repositories/users.

\paragraph{Implementation and Scalability.}
Given the large scale of user-project activity on GitHub, scalable computation is essential. We implement this process using distributed Spark-based pipelines, operating on user-project interaction logs stored in TiDB~\cite{tidb}. This allows efficient monthly recalculation of influence features across 115,466 repositories and millions of users.

\paragraph{Robustness and Expressiveness.}
This user-centric formulation offers several advantages. By taking into account the historical quality of user participation, it distinguishes habitual contributors from one-off actors and thus uncovers subtle signals of project health. Additionally, the graph-based design reduces susceptibility to manipulation, as artificially inflated surface features do not result in higher network-based influence. Further, the incorporation of temporal weighting highlights sustained or early engagement, which is closely associated with long-term project sustainability. Ultimately, this yields a more sturdy insight into project attention and maintenance status than what surface features can reveal.

From a program comprehension standpoint, these user influence signals directly support dependency vetting and ecosystem analysis by providing context on the quality of the user base for each project. Understanding which projects attract and retain influential users aids developers in evaluating the reliability and trustworthiness of their dependencies during comprehension, onboarding, or refactoring activities. 

\paragraph{Normalization for Temporal and Cross-Project Comparison.}

Our user-centric features exhibit a heavy-tailed distribution, typical in software engineering~\cite{goeminne2011pareto, 8862903}. To enable cross-project and temporal comparison, we apply two normalization strategies: (1) percentile ranking within each time window~\cite{9631870}, and (2) Z-score normalization on log-transformed values to mitigate skewness:

\begin{equation}
\begin{aligned}
    w_{\%} = \frac{rank(w_i)}{|i|} , \quad
    w_z = \frac{\ln w_i - \mu}{\sigma}
\end{aligned}
\end{equation}

These normalized features are used in downstream models, ensuring stability and interpretability for survival analysis and classification.

\subsubsection{Maintainer-Centric Features}
\label{subsubsec:maintainer-features}

Maintainers hold primary responsibility for project evolution, governance, and the formal declaration of abandonment; consequently, their behavioral signals serve as salient predictors of project risk. However, prior studies primarily rely on repository-level activity intervals (e.g., consecutive inactive days) derived from commit logs, which overlook this critical human factor~\cite{COELHO2020106274,DECAN2020110573}. Furthermore, some studies attempted role categorization without real community roles and did not achieve high C-index~\cite{10.1145/3236024.3236062}. To address these gaps, we propose a novel set of Maintainer-Centric Features that directly track maintainer engagement and responsiveness. Our approach moves beyond these limitations by first explicitly identifying the maintainers and then modeling their behavioral signals, providing a more direct and powerful predictor of abandonment risk.

In addition to their role in abandonment prediction, maintainer-centric features directly assist program comprehension tasks. For example, knowledge of maintainer activity intervals and response patterns enables developers to gauge project stewardship, identify potential bottlenecks in collaboration, and understand governance structures—key factors when making maintenance, onboarding, or technical debt management decisions. 

\paragraph{Maintainer Identification.}
We identify maintainers for each repository and time window by analyzing \emph{commit} history and \emph{PR} metadata. Specifically, a user is labeled as a maintainer if they either (1) perform direct \emph{commits}, or (2) merge \emph{PRs}. This operational criterion captures users with active governance roles, enabling accurate maintainer tracking over time.

\paragraph{Maintainer Activity Recency.}
An important feature of governance continuity is the time elapsed since the last maintainer activity. For each repository and timestamp, we compute the \emph{maintainer inactivity interval} as the duration since the most recent repository-level action performed by any identified maintainer. This inclusive approach captures the latest evidence of maintainer presence, beyond privileged operations like \emph{merges} or direct \emph{commits}. A prolonged inactivity interval strongly signals disengagement and is frequently a precursor to abandonment.

\paragraph{Maintainer Response Latency.}
Healthy open source projects typically exhibit prompt maintainer responses to externally submitted \emph{PRs} and \emph{issues}, indicating ongoing stewardship and community integration. As the risk of abandonment grows, response times elongate or become absent altogether. We capture this dynamic through two related features:
\begin{itemize}
    \item \textbf{Mean Response Time (Last 6 Months):} The average delay for maintainers to address new \emph{PRs} and \emph{issues} within the most recent 6-month window, quantifying timeliness of governance actions.
    \item \textbf{Change in Response Time (Last 6 Months):} The temporal trend, measured as the delta in mean response time over sequential intervals, providing early detection of deteriorating maintainer attention.
\end{itemize}

\subsubsection{Project Evolution Features}
\label{subsubsec:progress-features}

Beyond user and maintainer dynamics, the long-term evolution of an OSS project reveals critical signals of sustainability, community health, and risk of abandonment. To capture these deeper aspects, we construct a suite of project evolution features that build upon and extend prior concepts. For instance, while prior studies~\cite{10.1145/2901739.2903495,10.1145/3340544} have discussed the bug-fix ratio, we are the first to systematically incorporate it, along with the changing collaboration structure, development focus, and periodic behavioral patterns, into a predictive model for abandonment.

For program comprehension, these evolution features help developers and architects grasp the shifting focus, maturity, and collaborative structure of a project over time. By interpreting transitions in feature versus bug-fix focus, contributor diversity, and deviations from historical behavior, stakeholders can more accurately assess project trajectories, anticipate maintenance challenges, and plan sustainable architecture or integration strategies.

\paragraph{Community Participation Balance.}
Mature open source projects typically follow an ``onion model'' comprising core maintainers, active contributors, and peripheral users~\cite{onion}. Disruption to the balance among these roles---such as disproportionate maintainer dominance or a decline in contributor diversity---often precedes project abandonment. We operationalize this via:
\begin{itemize}
    \item \textbf{Maintainer Contribution Ratio ($p$)}: The proportion of key repository interactions (e.g., \emph{commits}, \emph{PR merges}) performed by maintainers in a recent window (e.g., six months).
    \item \textbf{User Activity Gini Coefficient ($G$)}:
    \begin{equation}
        G = \frac{\sum_{i=1}^{n} \sum_{j=1}^{n} |x_i - x_j|}{2n^2\mu}
    \end{equation}
    where $n$ is the number of users interacting with the project, $x_i$ the count of actions by user $i$, and $\mu$ the average number of actions. A high $G$ indicates concentration of activity and potential fragility in collaboration.
    \item \textbf{Participation Imbalance ($d$)}: The squared deviation from an ideal maintainer ratio: $d = (p - 0.5)^2$.
    Lower values indicate a healthy, balanced distribution of responsibilities, while high values signal imbalance and increased abandonment risk.
\end{itemize}

\paragraph{Development Focus via PR Categories.}
The relative focus of project development---whether on adding new features or on bug fixing---naturally shifts as a repository matures. A transition from innovation-driven contributions to predominantly bug-fixing is often indicates declining momentum:
\emph{PRs} are classified as \emph{feature-oriented} or \emph{bug-fix} based on curated keyword matching in PR titles (e.g., ``feature'', ``add'' for features; ``fix'', ``bug'' for bug-fixes).
For each time window, we extract: the ratio of feature to bug-fix \emph{PRs}, as well as the individual proportions of each category.
The shrinking feature/bug-fix \emph{PR} ratio is an early marker of functional stagnation and elevated risk of upcoming abandonment.

\paragraph{Deviation from Periodic Activity.}
Many OSS projects demonstrate regular, periodic activity patterns reflecting release planning or cyclical maintenance behaviors. Disruption or breakdown of these patterns
often signals the onset of decline:
\begin{itemize}
    \item \textbf{Activity Deviation:} For each repository, we compute the average of core activity features (e.g., \emph{commits}, \emph{issues}, \emph{PRs}) over the recent quarter/3 months and compare this to their historical baseline using z-score normalization.
    \item \textbf{Quarterly Deviation:} Activity in each recent quarter is standardized against the average of the previous three quarters, highlighting abrupt slowdowns or surges in development efforts.
\end{itemize}
A consistent pattern of negative deviation serves as a quantifiable warning signal preceding abandonment events.

\subsection{Feature Effectiveness Validation}
\label{subsec:feature-test}
Just as physiological indicators can predict health outcomes, behavioral traces of OSS repositories can serve as ``biomarkers'' for project vitality. In our setting, the interval from repository creation to abandonment constitutes the software ``lifespan.'' Survival analysis is well suited for this task, as it effectively handles time-to-event data with extensive right-censoring---matching the characteristics of our dataset. In this section, we evaluate the utility of our proposed features.

\paragraph{Suitability of Survival Analysis.}
Survival analysis is particularly apt for this problem because: (1) OSS projects may not all undergo abandonment within the observation window, resulting in right-censored data. (2) Survival models naturally support event-time analysis, enabling either risk ranking (``which projects are likely to be abandoned sooner?'') or direct estimation of abandonment time.

Rather than predicting a point estimate of abandonment time, survival models rank project risk according to concordance with observed lifespans. The concordance index (C-index) is therefore the primary evaluation metric, measuring the fraction of predicted risk orderings that match actual event orderings; it ranges from 0.5 (random guessing) to 1.0 (perfect discrimination). Empirically, a C-index above 0.8 is considered strong for prediction~\cite{HSPM}.

\paragraph{Accelerated Failure Time (AFT) Model.}
We instantiate the AFT approach using XGBoost’s survival module~\cite{Barnwal02102022,NEURIPS2021_7f6caf1f,PredictingIoT,10.1111/j.2517-6161.1972.tb00899.x}, where the log of event time (maintenance lifespan) is regressed on input features.

The AFT model supports efficient optimization with right censored samples using negative log-likelihood loss ($\texttt{nloglik}$). When applying the AFT model, it is standard practice to randomly split the dataset into training and testing sets~\cite{Barnwal02102022,NEURIPS2021_7f6caf1f,PredictingIoT}. In line with this convention, we randomly partitioned our data into 80\% for training and 20\% for testing, and employed early stopping during the model training to prevent overfitting.
We observe rapid convergence and stable loss minimization of the AFT (XGBoost-based) model in training.

\paragraph{Ablation Study and Predictive Contribution.}
To rigorously assess the marginal and joint contribution of different feature groups, we conduct ablation experiments using AFT models trained on various combinations. Table~\ref{tab:eval-model} reports the resulting Harrel’s and Uno’s C-index scores for the following feature sets:

\begin{itemize}
    \item \textbf{S}: Surface features (in Section~\ref{subsubsec:surface-features});
    \item \textbf{H}: HITS influence scores (excluding design in Section~\ref{subsubsec:weight});
    \item \textbf{U}: User-centric features (in Section~\ref{subsubsec:weight});
    \item \textbf{M}: Maintainer-centric features (in Section~\ref{subsubsec:maintainer-features});
    \item \textbf{P}: Project evolution features (in Section~\ref{subsubsec:progress-features}).
\end{itemize}

\begin{table}[ht]
    \centering
    \caption{C-index of AFT models under different features.}
    \vspace{-0.6em}
    \begin{tabular}{lcc}
        \toprule
        \textbf{Features} & \textbf{Harrel's C-index} & \textbf{Uno's C-index} \\
        \midrule
        S & 0.748 & 0.653 \\
        S $-$ stars & \underline{0.689} & \underline{0.621} \\
        S $+$ H & 0.810 & 0.716 \\
        S $+$ U & 0.826 & 0.751 \\
        S $+$ M & 0.778 & 0.670 \\
        S $+$ P & 0.767 & 0.665 \\
        U $+$ M $+$ P & 0.838 & 0.773 \\
        All & \textbf{0.846} & \textbf{0.781} \\
        \bottomrule
    \end{tabular}
    \vspace{-1em}
    \label{tab:eval-model}
\end{table}

We observe that augmenting the surface features (\textbf{S}) with any individual feature group consistently improves predictive performance. Notably, user-centric features (\textbf{U}) yield the largest gain, increasing Harrel's C-index from 0.748 to 0.826---surpassing the effect of adding HITS scores (\textbf{H}) alone (0.810). Furthermore, using only \textbf{U}, \textbf{M}, and \textbf{P}---excluding all surface features---achieves near-parity with the full model (0.838 vs. 0.846), demonstrating the sufficiency and complementarity of the proposed features.

\begin{figure}[ht]
    \centering
    \vspace{-0.5em}
    \includegraphics[width=1\linewidth]{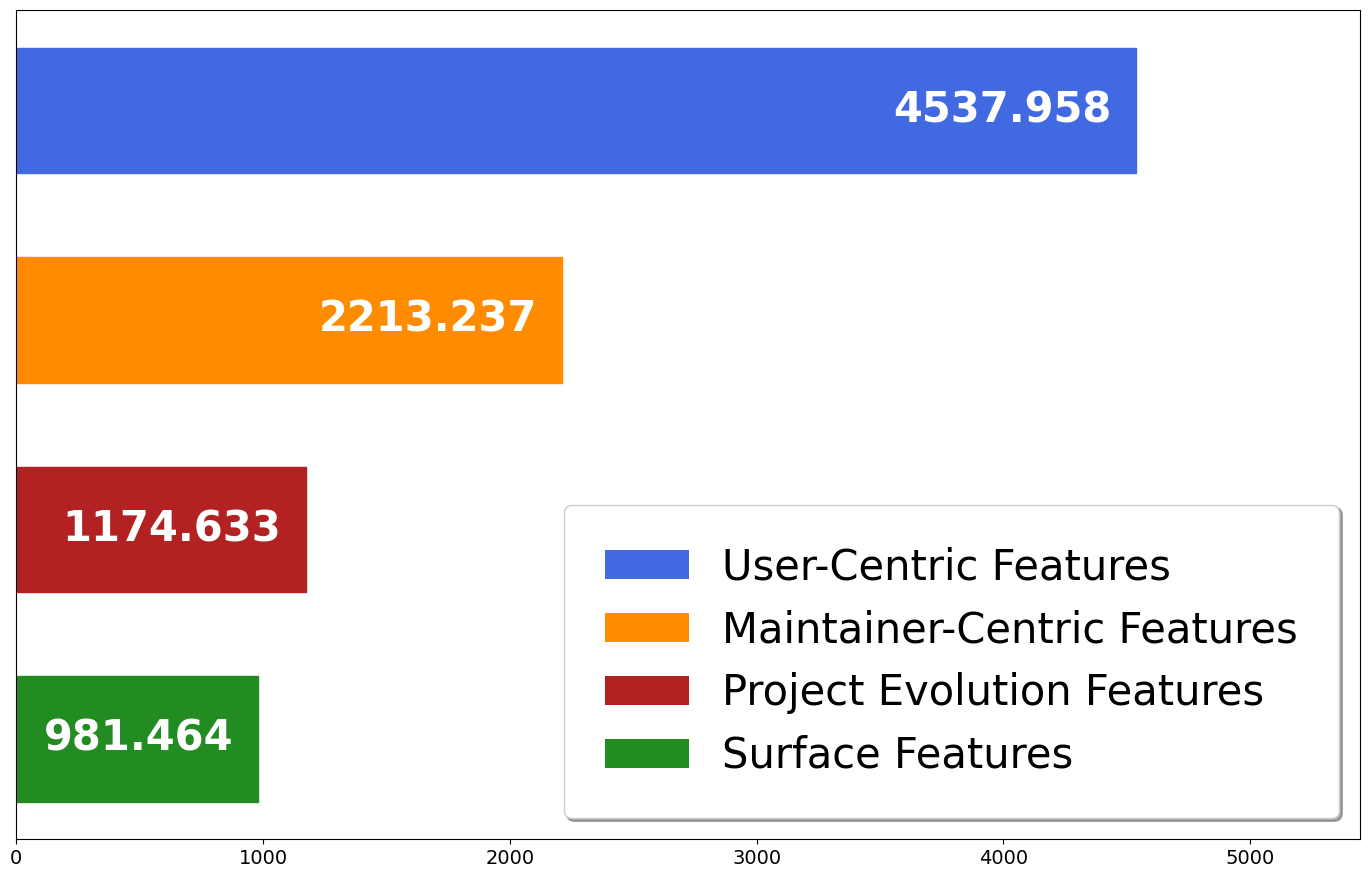}
    \vspace{-2em}
    \caption{Feature importance across categories.}
    \vspace{-0.5em}
    \label{fig:fiture-importance}
\end{figure}

\paragraph{Feature Importance via F-score Analysis.}
To elucidate the relative contribution of different feature groups, we leverage XGBoost's built-in F-score metric, which reflects how frequently a feature is used to split nodes across the ensemble~\cite{10.1145/2939672.2939785}. For each feature group, we aggregate the F-scores of its constituent features using a geometric mean (i.e., $\exp(\mathrm{mean}(\log(\text{F-score})))$), which smooths the influence of extreme values and provides a more balanced importance estimate across heterogeneous features. Figure~\ref{fig:fiture-importance} visualizes the aggregated scores for each group.

The results show that \textbf{User-Centric Features} contribute most prominently to the model, highlighting the predictive value of user engagement patterns~\cite{9796216}. \textbf{Maintainer-Centric Features} follow with a substantial score of 2213.24, suggesting the importance of maintainer activity signals. \textbf{Project Evolution Features} also exhibit meaningful predictive strength, capturing long-term development dynamics. In contrast, \textbf{Surface Features} contribute the least, indicating limited standalone utility in abandonment risk modeling.

\paragraph{Model-Agnostic Interpretability Using SHAP}
To comprehensively validate feature influence, we employ SHAP (SHapley Additive exPlanations)~\cite{10.5555/3295222.3295230}, a advanced method in explainable AI research~\cite{2021arXiv211001889B}. SHAP quantifies the marginal contribution of each feature to predicted risk scores by averaging over all possible feature combinations, ensuring mathematically consistent and fair importance attribution. SHAP's inherent robustness to feature collinearity ensures reliable importance attribution even when features exhibit moderate correlations.

The beeswarm SHAP visualization demonstrates that \emph{latest\_ maintainer\_activity\_interval}, \emph{weight\_zscore}, and \emph{contrib\_diversity} emerge as the three most influential predictors. These 3 features representing each of our novel feature categories, consistently exhibit the strongest directional impact on model predictions, confirming their critical role in abandonment predicting.

\paragraph{Case Illustration: Temporal Dynamics of Feature Signals.}

To illustrate the advantages and interpretability of our newly engineered features, we examine the time-series evolution of multiple signals from the \emph{Electronic WeChat} project~\cite{electronic-wechat}. All feature values are min-max normalized to the [0.1, 1] range for a beautiful view.

\begin{figure}[t]
    \centering
    \includegraphics[width=1\linewidth]{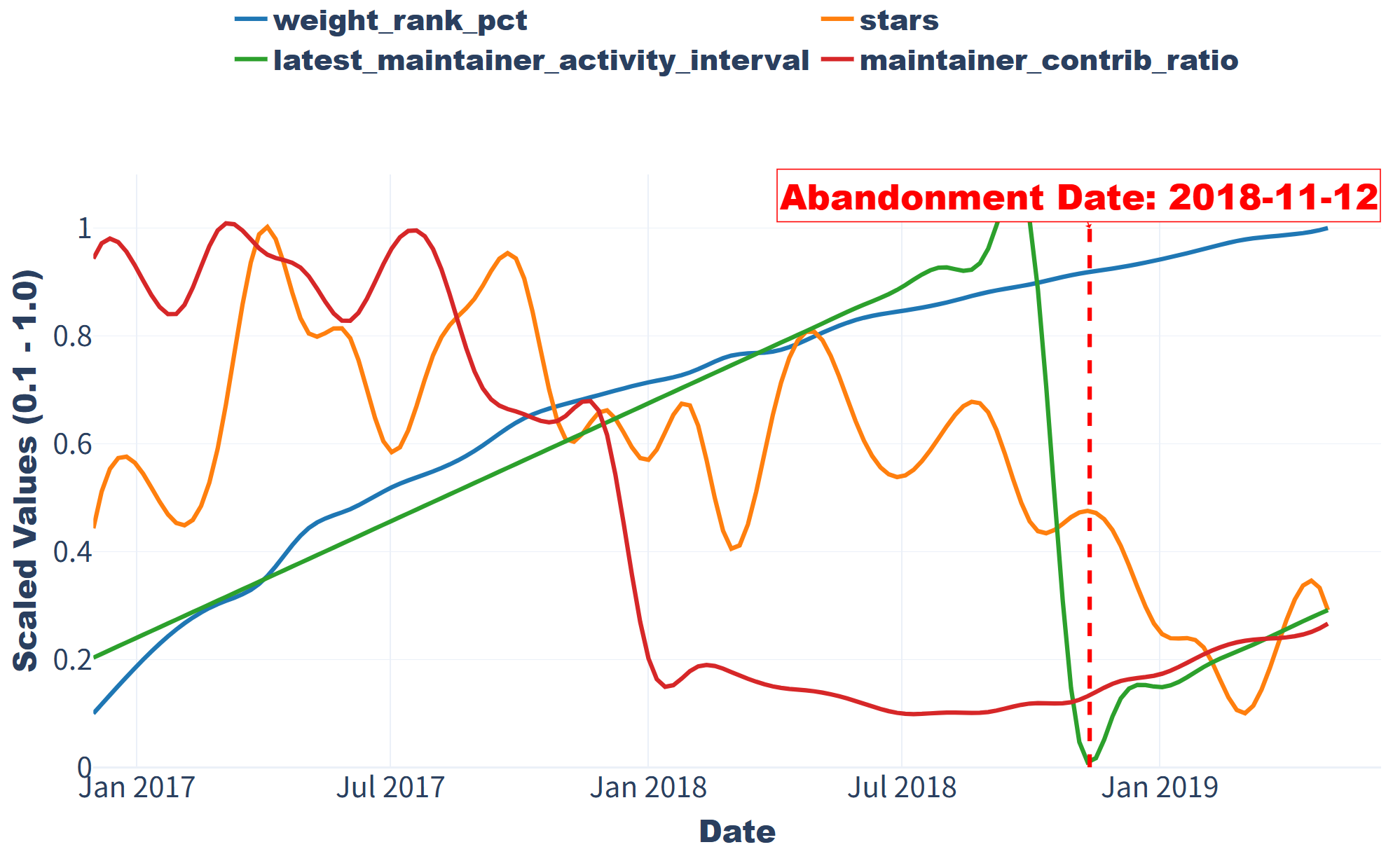}
    \vspace{-2em}
    \caption{Time-series feature dynamics for \emph{Electronic WeChat}.}
    \vspace{-1em}
    \label{fig:wechat-features}
\end{figure}

This project, a popular third-party MacOS WeChat client, was abandoned after Tencent officially released WeChat for Mac 2.0 on August 16, 2016, which has been updating regularly ever since.

From Figure~\ref{fig:wechat-features}, we observe:

\textbf{Surface Feature (Orange):}
The number of stars remained consistently high before the abandonment event, reflecting the project's lasting popularity. However, this persistent star growth results mainly from historical reputation and user recommendations, making it insensitive to changes in actual maintenance or activity. This observation aligns with our earlier discussion in Section~\ref{subsubsec:surface-features}, confirming that the inherent limitations of surface metrics like star count, limit their effectiveness as a timely warning signal for project abandonment.

\textbf{User-centric Feature (Blue):}
The percentile rank of interaction weight exhibits a steady increase (indicating a decline in relative ranking) as abandonment approaches. This change indicates a decline in genuine community engagement, even while stars remain high. As fewer influential users interact meaningfully with the project, this feature effectively signals a loss of capable user support, often preceding maintenance termination.

\textbf{Maintainer-centric Feature (Green):}
The time since the last maintainer activity increases continuously prior to abandonment, confirming a complete cessation of involvement. This extended inactivity interval serves as a quantitative indicator that the project has been neglected by its maintainers, directly foreshadowing the eventual abandonment event.

\textbf{Project Evolution Feature (Red):}
Initially, the maintainer contribution ratio is high but then drops sharply before abandonment. This shift reflects a breakdown in the stable collaboration pattern, signaling impending inactivity.

This case underscores the critical need for advanced, multi-perspective features---beyond traditional surface features---when predicting OSS abandonment with precision and interpretability. Our three newly proposed feature types capture a broader range of repository state information than surface activity features.
Such case studies demonstrate how our multi-perspective feature set can offer not only predictive alerts but also actionable narratives for project stakeholders to interpret and comprehend the abandonment process in real time. 

\paragraph{Comprehensive Feature Efficacy}
Collectively, the ablation studies, F-Score analysis, SHAP, and case analysis provide converging evidence that all feature categories summarized in Table~\ref{tab:feature-categories} contribute effectively to abandonment prediction. 
Together, these multi-perspective features improve prediction accuracy and give developers interpretable signals for understanding project health, collaboration, and evolution, supporting both risk assessment and program comprehension. 

\section{Real-World Deployment}
\label{sec:framework-application}

In this section, we evaluate the practical utility of our prediction method by deploying it within the \textbf{openEuler} ecosystem, a production-grade, community-driven Linux distribution, demonstrating methodological generalizability and real-world impact.

\subsection{Practical Risk Prediction via GBSA Classifier}
\label{subsec:gbsac}

To address the needs of stakeholders requiring binary, near-term risk assessments---for example, ``Will this package likely cease maintenance within the next 6 months?''---we employ a \emph{Gradient Boosted Survival Analysis} (GBSA) classifier using the feature set presented in Section~\ref{sec:feature-framework}. Unlike traditional regression-based survival models, GBSA excels at handling high-dimensional, time-varying, and right-censored datasets in a classification setting, and is well suited for integration into real-time package quality monitoring systems.

\paragraph{Formulation.}
For deployment, we treat abandonment prediction as a temporal binary classification problem. Let $T$ denote a reference date (in this study, July 1, 2018), and $\Delta t$ the prediction horizon (set to 6 months). For each repository, we aggregate all features up to $T$, and label as ``positive'' those repositories for which unambiguous abandonment occurs between $T$ and $T + \Delta t$. Repositories not ceasing maintenance in this window are used as ``negative'' or censored samples.
We select July 1, 2018 as the reference date $T$ because a relatively large number of openEuler ecosystem packages experienced abandonment during this period of time. This maximizes the number and diversity of positive samples for robust model evaluation.

\paragraph{Training and Evaluation.}
We balance positive and negative samples to address class imbalance and apply Bayesian optimization (via Optuna~\cite{optuna_2019}) for hyperparameter tuning (e.g., learning rate, tree depth, subsample ratio). Model performance is evaluated using accuracy, precision, recall, and the Harrell’s C-index, consistent with operational needs.

Our GBSA classifier achieves an overall prediction accuracy of 78.91\% on an unbalanced test set, confirming that our engineered features and survival-based modeling retain high discriminative power in a practical early warning context.

\subsection{Deployment in the openEuler Ecosystem}
\label{subsec:application-openeuler}

openEuler~\cite{openeuler,openeuler-url} is a widely used open-source Linux OS serving critical infrastructure, with over 5 million new installations in 2024~\cite{openEuler-1000}. This scale makes proactive package abandonment risk monitoring essential.

\subsubsection{Mapping openEuler Packages to Upstream GitHub Repositories}
\label{subsubsec:oe-dataset}

openEuler is an open-source operating system whose source code is primarily hosted on Gitee~\cite{gitee}, comprising two key repositories: the code repository for upstream source projects~\cite{gitee-openEuler}, and the package repository for build-ready software packages used in actual releases~\cite{gitee-src-openEuler}. These packages are directly tied to system distributions and user security. 
To assess ecosystem-level abandonment risks, we take the openEuler release as a representative case and systematically map all its packages to their corresponding upstream GitHub repositories. Package-level metadata is retrieved from the official repository index~\cite{oe-package}.

\paragraph{Automated Mapping Pipeline.}
We design a seven-stage pipeline to associate openEuler-22.03-LTS \textsf{aarch64/loongarch64/x86\_64} packages with their corresponding upstream GitHub repositories:
\textbf{(1) Rule-based mapping:} Manually defined rules for common packages (e.g., \emph{texlive});
\textbf{(2) Gitee metadata:} Extracting GitHub URLs from the \emph{url} field in \emph{primary.xml};
\textbf{(3) Spec file parsing:} Locating source repositories and extracting \emph{Source0} from \emph{.spec} files via \emph{depchase}~\cite{depchase};
\textbf{(4) YAML field matching:} Parsing the \emph{git\_url} or \emph{src\_repo} fields with typo tolerance;
\textbf{(5) URL-based inference:} Inferring likely GitHub repos from homepage domains;
\textbf{(6) GitHub API search:} Fallback search by name, ranking by stars ($\geq$32);
\textbf{(7) Homepage crawling:} Scraping official websites to discover embedded GitHub links.

This automated pipeline successfully maps 14,284 out of 16,888 openEuler packages (coverage: 84.6\%) to valid GitHub repositories, as show in Table~\ref{tab:oe-distribution}. The resulting mapping forms the foundation for downstream risk inference.

\begin{table}[b]
    \centering
    \vspace{-0.7em}
    \caption{Distribution of mapping methods.}
    \vspace{-0.7em}
    \begin{tabular}{lc}
    \toprule
    \textbf{Method} & \textbf{Proportion}\\
    \midrule
    Rule-based & 59.32\% \\
    Gitee metadata & 20.4\% \\
    Spec & 13.7\%\\
    YAML & 7.7\% \\
    URL-based &  2.9\%\\
    API search &   11.4\%\\
    Homepage crawling & 2.4\%\\
    \bottomrule
    \end{tabular}
    \vspace{-1em}
    \label{tab:oe-distribution}
\end{table}

\subsubsection{Abandonment Prediction: Experimental Protocol}
\label{subsubsec:oe-predict}

For each mapped upstream repository, we extract all temporal features in Section~\ref{subsec:feature-design}, up to the reference time $T$, and apply the trained GBSA classifier to estimate the probability of abandonment within the next 6 months.

\paragraph{Evaluation Setting.}
We anchor prediction at $T = \text{July 1, 2018}$ and label abandonment outcomes occurring before January 1, 2019. Of the mapped repositories, 278 are included in our annotated dataset and have sufficient history and feature coverage for risk scoring. Of these, 7 repositories were actually abandoned between July 1, 2018, and January 1, 2019, forming the primary evaluation cohort.

\paragraph{Results.}
Our deployed GBSA model successfully flags 3 out of 7 repositories that were actually abandoned within the package set of openEuler, based on a reference date of July 1, 2018 and a 6-month prediction window. Accuracy, precision, and C-index achieve 0.9857, 1.0, and 0.8049, respectively.
The deployment within the openEuler ecosystem further showcases how actionable and interpretable risk signals foster improved program comprehension and risk awareness throughout large-scale software supply chains.

\paragraph{Interpretation.}
A manual review of the four missed cases (false negatives) shows that these repositories lacked the typical warning signals captured by our features prior to abandonment; most were abandoned abruptly or without clear early indicators. This highlights the limitation of behavioral features in detecting entirely unannounced or exogenous abandonment events.
Nevertheless, our model achieves strong overall accuracy and precision, reliably identifying early risk in the vast majority of practical cases, which supports deployment in real-world package management scenarios.

\section{Threats to Validity}

\paragraph{Internal Validity.}
First, our definition of abandonment, relying on explicit archival status or semantic cues in documentation, may still incompletely capture all true abandonment events. Silent abandonments, especially among small or low-profile projects, could be missed. Additionally, although our feature engineering seeks to comprehensively model user, maintainer, and project evolution dynamics, certain abrupt or exogenous abandonment events (e.g., organizational, legal, or funding shocks) may not manifest as detectable signals in repository behavior or metadata prior to abandonment, potentially resulting in false negatives as observed in some openEuler cases (see Section~\ref{subsec:application-openeuler}).

\paragraph{External Validity.}
Our study primarily targets GitHub-hosted repositories, which, while representing the majority of global OSS activity, may not generalize to other hosting platforms such as Gitee, GitLab, or self-hosted infrastructures. These alternatives may differ in developer conventions, community dynamics, and the richness or accessibility of behavioral data.

\section{Conclusion}

OSS project abandonment poses significant risks to software supply chains and ecosystem stability. This work introduces a precise, actionable discrimination method for abandonment, effectively addressing the low efficiency and high false-positive rates prevalent in prior studies.
We construct a comprehensive longitudinal dataset to support robust empirical analysis and propose a multi-perspective feature framework that integrates user influence, maintainer activity, and project evolution indicators. Our approach significantly outperforms traditional surface features, achieving a C-index of 0.846 through survival analysis.
Finally, deployment within the openEuler ecosystem demonstrates the practical utility of our approach for proactive risk identification in real-world package management scenarios.

As dependency on OSS continues to increase, scalable, early-warning tools such as ours are essential for ensuring the sustainability and security of the software ecosystem.
Our interpretable framework supports program comprehension by providing actionable risk signals for dependency vetting, onboarding, and architectural planning, thus helping ensure sustainable OSS evolution. 

\bibliographystyle{ACM-Reference-Format}
\bibliography{ref}

\end{document}